\documentstyle[preprint,aps]{revtex}
\begin{document}
\hsize = 7.0in
\tighten
\preprint{\bf Published in Phys. Rev. D  47 (1993) 5161-5168}

\widetext
\def\eq{\,=\,}
\def\beq{\begin{equation}}
\def\eeq{\end{equation}}
\def\l{\left(}
\def\r{\right)}
\def\po{\mathaccent 23 p}
\def\pp{p^+}
\def\pr{p_r}
\def\pl{p_\ell}
\draft
\title{
Front Form Spinors in  Weinberg-Soper Formalism
and Melosh Transformations for any Spin\footnotemark
\footnotetext 
{The work of DVA
 was done under the auspices of the U. S. Department of Energy.
The work of MS was supported by
the U. S. National Science Foundation
 under the Grant PHY-8907852.
}}
\author{D. V. Ahluwalia \dag \ and Mikolaj Sawicki \ddag}
\address{
\dag Medium Energy Physics Division, MS H-844, Los Alamos
National Laboratory\\
 Los Alamos, New Mexico 87545, USA.}
\address{
\ddag Department of Physics, Iowa State University, Ames, Iowa 50011, USA.
}
\maketitle
\begin{abstract}
Using the Weinberg-Soper formalism we construct the front form
$(j,0)\oplus(0,j)$ spinors. Explicit expressions for the generalised Melosh
transformations up to spin two are obtained. The formalism, without explicitly
invoking any wave equations, reproduces spin one half front-form results of
Melosh, Lepage and Brodsky, and Dziembowski.
\end{abstract}

\section{Introduction}
\label{sec:intro}

Following the work of Weinberg \cite{Weinberg} and Soper \cite{Soper}
we extend our recent work \cite{earlier} and explicitly construct the
front form covariant spinors for arbitrary spin $j$ {\it without} a direct
reference to any wave equation. We also obtain a  generalisation of Melosh
transformation, $\Omega(j)$, to $(j,0)\oplus(0,j)$ fields.
Explicit examples for $j\,=\,{1\over 2},\,1,\, {3\over 2},
\,2$ are given. For $j\,=\,{1\over 2}$, the formalism reproduces the
front form spinors of Lepage and Brodsky
\cite{lepage}; and $\Omega({1\over 2})$ coincides with the celebrated ``Melosh
transformation'' \cite{Melosh,Dziembowski}. This approach is in the spirit of
Dirac's \cite{DiracSpin} original motivation for the high-spin wave equations
for ``approximate application to composite particles'' and opens up the
possibility of  constructing a QCD based effective field theory of hadronic
resonances along the lines of recent work of Cahill \cite{Cahill}.

Unless otherwise indicated we follow the notation of Refs. [9, 3(b)].
We use the notation $x^\mu\eq\left(x^+,\,x^1,\,x^2,\,x^-\right)$. In terms of
the instant form variables $\underline
x^\mu\eq\left(x^\circ,\,x^1,\,x^2,\,x^3\right)$, we have $x^\pm\eq
x^\circ\,\pm\,x^3$. The evolution of a system is studied along the coordinate
$x^+$, and as such it plays the role of ``time.''
 
\section{FRONT FORM HADRONIC SPINORS}
 
We will work under the assumption that the center of mass of a composite
hadrons is best described, in the phenomenological sense, by the fields
constructed from the $(j,0)\oplus(0,j)$ spinors in the front form. Following
the Weinberg-Soper formalism \cite{Weinberg,Soper}, 
these spinors will be constructed from the right handed,
$\phi^{R}(p^\mu)$, and left handed, $\phi^{L}(p^\mu)$, matter fields.
We begin  from the transformation which takes a particle
from rest, $\po^\mu\eq(m,\,0,\,0,\,m)$,
to a particle moving with an arbitrary four momentum $p^\mu\eq
\left(p^+,\,p^1,\,p^2,\,p^-\right)$
[Note: for massive particles $p^+ \,>\, 0$.].
 
In the {\it instant form of field theory} the transformation 
which takes the rest momentum
${\underline\po}^\mu\eq \left(m,\,0,\,0,\,0\right)\,\to\,
\underline {p}^\mu\eq(p^\circ,\,\vec p\,)$, 
is constructed out of the boost operator $\vec K$ and is given by
\beq
{\underline p}^\mu \eq{\Lambda^\mu}_\nu\,\underline\po^\nu\,,
 \label{i}
\eeq
with
\beq
\Lambda\eq\exp\left(i\vec \varphi\cdot\vec K\right)\,\,.\label{il} \eeq
In Eq. (\ref{il}) the boost parameter $\vec \varphi$ is defined as follows
\beq
\hat\varphi\eq{\vec p/|\vec p\,|},~~\cosh(\varphi)\eq E/m,~~\sinh(\varphi)\eq
|\vec p\,|/m.\label{ibp}
\eeq
Note that the stability group of the $x^\circ\eq 0$ plane consists of the six
generators $\vec J$ and $\vec P$. The $\vec K$, along with the $P^\circ$,
generate the instant-form dynamics. 
 
In the {\it front form of field theory} the transformation which takes
${\underline\po}^\mu\eq \left(m,\,0,\,0,\,0\right)\,\to\,
\underline p^\mu\eq(p^\circ,\,\vec p\,)$
is defined  \cite{Soper,LeutwylerS}  by
\beq
\underline{p}^\mu\eq {L^\mu}_\nu\,\underline{\po}^\nu, \label{onee}
\eeq
with the matrix $L$ given by
\beq
L\eq\exp\left(i\,\vec v_\perp\cdot\vec G_\perp\right)
\,\exp\left(i\,\eta K_3\right). \label{boost}
\eeq

The parameters  $\eta$ and $\vec v_\perp = (v_x, v_y)$ specify a given boost. 
The  generators $\vec G_\perp$
are defined as follows
\beq
G_1\eq K_1\,-\,J_2,~~G_2\eq K_2\,+\,J_1,\label{g}
\eeq
and together with
\beq
P_- \eq P_\circ\, -\, P_3,~~P_1,~~P_2,~~J_3,~~K_3\quad,
\eeq
form the seven generators of the stability group of the
 $x^+\eq 0$ plane. (Note that $P_- = P^+$).
The algebra associated with the stability group is summarised in Table I. 
The generators $D_1\eq K_1\,+\,J_2$, $D_2\eq K_2\,-\,J_1$ and $P_+\eq
P_\circ \,+\, P_3$ generate the front-form dynamics.

It is important to note that while the front-form 
transformation $L$ is specified
entirely in terms of the generators of the $x^+\eq 0$ plane stability group, the
instant-form transformation $\Lambda$ involves dynamical generators associated
with the $x^\circ\eq 0$ plane.

Using the explicit matrix expressions for $\vec J = (J_{1},J_{2},J_{3})$ and $\vec K = (K_{1},K_{2},K_{3})$ given in
Eqs. (2.65 - 2.67) of Ref.
\cite{Ryder} we obtain an explicit expression for
the  boost $L$ defined in Eq. (\ref{boost})
\widetext
\begin{eqnarray}
[{L^\mu}_\nu]\eq
\left[
\begin{array}{ccccccc}
\cosh(\eta)+{1\over 2} \vec v_\perp^{\,2} \exp(\eta)&\; & v_x &\;& v_y &\;&
\sinh(\eta)+{1\over 2} \vec v_\perp^{\,2} \exp(\eta)\\
v_x \exp(\eta)&\; & 1 &\;& 0 &\;&v_x \exp(\eta)\\
v_y \exp(\eta) &\;& 0 &\;& 1 &\;&v_y \exp(\eta)\\
\sinh(\eta)-{1\over 2} \vec v_\perp^{\,2} \exp(\eta) &\;& v_x &\;& v_y &\;&
\cosh(\eta)-{1\over 2} \vec v_\perp^{\,2} \exp(\eta)\\
\end{array}
\right]
. \label{l}
\end{eqnarray}

Recalling that the components of the front form momentum $p^\mu$ are defined as
$p^\pm\eq p^\circ\,\pm\,p^3$ this yields 

\widetext
\beq
p^+\eq m\,\exp(\eta),~~\vec p_\perp\eq m\,\exp(\eta)\,\vec v_\perp,~~p^-
\eq m\,\exp(-\eta)\,+\,m\,\exp(\eta) \,\vec v_\perp^{\,2}\,\,.\label{pp}
\eeq

The variables $\eta$ and  $\vec v_\perp$  are fixed by requiring
$\underline {p}^\mu$ generated by Eq. (1) to be identical to the $\underline
{p}^\mu$ produced by Eq. (4).
 
Given the transformation $L$, Eq. (\ref{boost}), we now wish to construct
the front form
$(j,0)\oplus(0,j)$ hadronic spinors. To proceed in this direction we
rewrite $L$ by expanding the exponentials in Eq. (\ref{boost}),
and using Table I to arrive at \cite{msBHF}
\beq
L\eq\exp\left[i\left(a\, \vec v_\perp\cdot\vec G_\perp\,+\,\eta\,
 K_3\right)\right],
 \label{mikolaj}
\eeq
with
\beq
a\eq{\eta\over{1\,-\,\exp(-\eta)}}.\label{mikolaja}
\eeq
For the $(j,0)$ matter fields, $\phi^R(p^\mu)$, we have
[1,9,3(g)]  $\vec K\eq -\,i\,\vec J$. For the $(0,j)$ matter
fields, $\phi^L(p^\mu)$,  $\vec K\eq +\,i\,\vec J$.  Using this observation,
along with Eq. (\ref{mikolaj}) and definitions (\ref{g}), we obtain the
transformation properties of the front form $(j,0)$ and $(0,j)$ hadronic fields
\beq
\phi^R(p^\mu)\eq\exp\left(+\,\eta \,\hat b\cdot\vec J\,\right)\,\phi^R(\po^\mu),
\label{right}
\eeq
and
\beq
\phi^L(p^\mu)\eq\exp\left(-\,\eta\, \hat b^\ast\cdot\vec J\,\right)\,
\phi^L(\po^\mu),
\label{left}
\eeq
where the  unit vectors $\hat b$
and $\hat b^\ast$ are given by
\begin{eqnarray}
\hat b&\eq&\eta^{-1}\, \left(a\,v_r\,,\,\,-i\, a\, v_r\,,\,\,\eta\right)\, ,
\label{b}
\\
\hat b^\ast&\eq& \eta^{-1}\,\left(a\,v_\ell\,,\,\,i\, a\, v_\ell\,,\,\,
\eta\right)\,;~~~~~~~~\hat b\cdot
\hat b\eq 1\eq \hat b^\ast\cdot\hat b^\ast\,,\label{bs}
\end{eqnarray}
with $v_r\,\eq\,v_x\,+\,i\,v_y$ and
$v_\ell\,\eq\,v_x\,-\,i\,v_y$.
 
We now make two observations. First,  under the operation of {\it parity}
we have $\phi^R(p^\mu)\,\leftrightarrow\, \phi^L(p^\mu).$
Second, for the particle at rest ($\vec p\eq\vec 0,\;\ p^\mu\eq\po^\mu$)
the concept of handedness looses its physical significance, (c.f.
Ref. \cite{Ryder} and more detailed discussion in Ref. \cite{prcb}),
and this in turn yields  the relation

\beq
\phi^R(\po^\mu)\eq\pm\,\phi^L(\po^\mu).\label{pm}
\eeq
We now introduce the spin-j hadronic spinor
\beq
\psi(p^\mu)\eq {1\over\sqrt{2}}
\left[
\begin{array}{c}
\phi^R(p^\mu)\,+\,\phi^L(p^\mu)\\ \\
\phi^R(p^\mu)\,-\,\phi^L(p^\mu)
\end{array}
\right],\label{psi}
\eeq
and observe that the  plus (minus) sign in Eq. (\ref{pm}) yields spinors with {\it even (odd)
spinor parity}. We will denote the even spinor parity spinors  by ${\cal
U}(p^\mu)$; and the odd spinor  parity spinors by ${\cal V}(p^\mu)$. 
 
The transformation property  for these hadronic spinors under the boost
(\ref{boost}) is now readily obtained by using  Eqs. (\ref{right}) and
(\ref{left}). The result is
\beq
\psi(p^\mu)\eq M(L)\,\,\psi(\po^\mu)\,,\label{tran}
\eeq
with the operator $M(L)$  given by
\widetext
\beq
M(L)\eq {1\over 2}\,\left[
\begin{array}{ccc}
\exp\left(\eta\,\hat b\cdot\vec J\,\right)
\,+\, \exp\left(-\,\eta\,\hat b^\ast\cdot\vec J\,\right)&~~~~~~~&
\exp\left(\eta\,\hat b\cdot\vec J\,\right)
\,-\, \exp\left(-\,\eta\,\hat b^\ast\cdot\vec J\,\right)\\ \\
\exp\left(\eta\,\hat b\cdot\vec J\,\right)
\,-\, \exp\left(-\,\eta\,\hat b^\ast\cdot\vec J\,\right)&~~~~~~~&
\exp\left(\eta\,\hat b\cdot\vec J\,\right)
\,+\, \exp\left(-\,\eta\,\hat b^\ast\cdot\vec J\,\right)
\end{array}
\right].\label{ml}
\eeq

In what follows we present the explicit construction of the hadronic
spinors ${\cal U}(p^\mu)$ and ${\cal V}(p^\mu)$ for $j\eq{1\over
2},\,1,{3\over 2},\,2$.

\section{Construction of ${\cal U}(\lowercase {p}^\mu)$ and ${\cal V}
(\lowercase{p}^\mu)$ and Their Properties}

Let us first note that the front form helicity operator
\beq
{\cal J}_3 \equiv J_3 \,+\,{1\over P_-}\left(G_1\,P_2\,-\,G_2\,P_1\right),
\label{o}
\eeq
introduced by Soper \cite{Soper} and discussed by Leutwyler
and Stern \cite{LeutwylerS}, commutes with all generators of the stability group
associated with the $x^+\eq0$ plane. The front form helicity  operator
associated with the $(j,0)\oplus(0,j)$ spinors constructed above is
then readily defined to be
\beq
\Theta\eq\left[
\begin{array}{ccc}
{\cal J}_3 &~~& 0\\
0 &~~&{\cal J}_3
\end{array}\right]\,\,. \label{th}
\eeq

If we choose a matrix representation of the $\vec J$ operators
with $J_3$ diagonal (we follow the standard convention of Ref.
\cite{Schiff}\ ) then  the $2(2j\,+\,1)$--element basis
spinors for a particle at rest have the general form

\widetext
\beq
{\cal U}_{+j}(\po^\mu)\eq
\left[
\begin{array}{c}
{\cal N}(j)\\
0\\
0\\
\vdots\\
0
\end{array}
\right],~~
{\cal U}_{j-1}(\po^\mu)\eq
\left[
\begin{array}{c}
0\\
{\cal N}(j)\\
0\\
\vdots\\
0
\end{array}
\right],\ldots\ldots,\,
{\cal V}_{-j}(\po^\mu)\eq
\left[
\begin{array}{c}
0\\
0\\
0\\
\vdots\\
{\cal N}(j)
\end{array}
\right]\,\,.\label{basis}
\eeq

The index $h\eq j,\, j-1,\,\ldots,\,-j$ on the ${\cal U}_h(p^\mu)$ and
${\cal V}_h(p^\mu)$ corresponds to the eigenvalues of  the front
form helicity operator $\Theta$. The spin-dependent normalisation
constant ${\cal N}(j)$ is to be so chosen that for the {\it massless}
particles all ${\cal U}_h(\po^\mu)$ and ${\cal V}_h(\po^\mu)$ vanish
(There can be no massless particles at rest !), and the only non-vanishing
spinors are ${\cal U}_{h\eq\pm j}(p^\mu)$ and ${\cal V}_{h\eq\pm
j}(p^\mu)$.
The simplest choice satisfying these requirements is
\beq
{\cal N}(j)\eq m^{\,j}.\label{nj}\label{norm}
\eeq
 
We now have all the details needed to construct ${\cal U}_h(p^\mu)$
and ${\cal V}_h(p^\mu)$ for any hadronic field. In this paragraph we
summarise the algebraic construction used for $j\eq{1\over
2},\,1,{3\over 2},\,2$. Using Eqs. (A27,A28) and (A31,A32) of Ref.
\cite{Weinberg} along with Eqs.  (\ref{pp},\ref{b},\ref{bs}) above we
obtain  the expansions for the $\exp\left(\eta\,\hat b\cdot\vec
J\,\right)$ and $ \exp\left(-\,\eta\,\hat b^\ast\cdot\vec J\,\right)$
which appear in the light-front 
$(j,0)\oplus(0,j)$-spinor boost matrix $M(L)$, Eq.
(\ref{ml}). These expansions are presented in Appendix A.  Using the
results of Appendix A, explicit expressions for $M(L)$, Eq.
(\ref{ml}), are then calculated as a simple, but somewhat lengthy
algebraic exercise. These expressions for $M(L)$ combined with Eqs.
(\ref{tran}) and (\ref{basis}) yield the hadronic spinors
presented in Appendix B. The generality of the procedure for any spin
is now obvious, and the procedure reduces to the well defined algebraic
manipulations.
 
We now introduce the following useful matrices,
\beq
\Gamma^\circ\eq
\left[
\begin{array}{ccc}
I&{~}&0\\
0&{~}&-I
\end{array}
\right]\,,
\Gamma^{5}\eq
\left[
\begin{array}{ccc}
0&{~~}&I \\ 
I&{~~}& 0
\end{array}
\right]\, ,
\label{go}
\eeq
with $I\eq (2j\,+\,1)\times(2j\,+\,1)$ identity matrix.

In reference to the  spinors presented in Appendix B, we define
\beq
\overline\psi\l p^\mu \r \equiv \psi^\dagger\l p^\mu\r\,\Gamma^\circ\, .
\label{psibar}
\eeq

Using the explicit
expressions for ${\cal U}(p^\mu)$ and ${\cal V}(p^\mu)$, Eqs.
(B1-B7), we verify that
\begin{eqnarray}
{\overline {\cal U}}_h (p^\mu)\,\,{\cal U}_{h'}(p^\mu) &\eq&
m^{2\,j}\,\delta_{h h'} \,,\label{three}\\
{\overline {\cal V}}_h (p^\mu)\,\,{\cal V}_{h'}(p^\mu) &\eq&
-\,m^{2\,j}\,\delta_{h h'} \,,\label{four}\\
{\overline {\cal U}}_h (p^\mu)\,\,{\cal V}_{h'}(p^\mu) &\eq& 0
\eq{\overline {\cal V}}_h (p^\mu)\,\,{\cal U}_{h'}(p^\mu)  \,\,.\label{five}
\end{eqnarray}

For spin one half the result given by Eq. (B1) corresponds to that given by
Lepage and Brodsky [4, Eq.A3]
\footnotemark \footnotetext{
   Note, however, a slightly different normalization and convention 
   chosen by Lepage and Brodsky for the odd spinor parity spinors:
   $
u_{\uparrow} = \sqrt{2} \,{\cal U}_{1/2}\,,\,
    u_{\downarrow} = \sqrt{2}\, {\cal U}_{-1/2}\,,\,
    v_{\uparrow} = \sqrt{2} \,{\cal V}_{-1/2}\,,\, 
    v_{\downarrow} = - \sqrt{2}\, {\cal V}_{1/2}
$. }.
A noteworthy feature of the
front-form spinors constructed here in the $(j,\,0)\oplus(0,\,j)$
representation is the observation that: 
\begin{enumerate}
\item 
For the massless case,
$m \eq 0$, the  ${\cal U}_h(p^\mu)$ and ${\cal V}_h(p^\mu)$  identically vanish
unless the associated front form helicity $h\eq\pm\,j$; and 
\item 
For
the massive case, $m \neq 0$, an examination of Eqs. (B3), (B4) and (B5-B7)
yields the result that in the ``high momentum'' limit, $p^+\gg\,m$, the
leading asymptotic behavior of ${\cal U}_h(p^\mu)$, and ${\cal V}_h(p^\mu)$ is
given by $\sim\,\l\pp\r^{|h|}$. 
\end{enumerate}
\noindent
The correspondence between observations I and II is to be perhaps fully
realised when various matrix elements, of appropriate front-form operators
$\cal O$,\quad $\overline\psi_{h'}(p'{^\mu})\,{\cal O}\,\psi_h(p^\mu)$, are
studied for the massive case. Since the high momentum behaviour of 
 ${\cal U}_h(p^\mu)$ and ${\cal
V}_h(p^\mu)$ is $\sim\,\l\pp\r^{|h|}$, we can expect to generalise to arbitrary
spin the results on helicity amplitudes presented by Lepage and Brodsky
\cite{lepage} for $j=1/2$. 
It is expected that the dominant {\it elastic-scattering}
amplitudes will correspond to the helicity non-changing processes,  while the
helicity changing transitions will be  suppressed by appropriate powers of the
factor $(m/{p^+})$. The fact that the front form spinors  ${\cal
U}_{|h|<j}(p^\mu)$ and ${\cal V}_{|h|<j}(p^\mu)$ for massive particles do {\it
not} identically vanish in the high momentum limit is of profound physical
significance. To see this note that , as argued by Brodsky and Lepage
\cite{sjb}, the hadrons in $e^+\,+\,e^-\, \rightarrow\,
\gamma^\ast\,\rightarrow\,{\cal H}_A\,+\,\overline{\cal H}_B$ are produced at
large $Q^2$ with opposite helicity $h_A\,+\,h_B\eq 0$ and $|h_i| \le 1/2$. As a
consequence, to give an example [14, Table I], the process $e^+\,+\,e^-\,
\rightarrow\, p_{\pm{1\over 2}} \,+\,\overline\Delta_{\pm{3\over 2}}$ is {\it
suppressed} relative to $e^+\,+\,e^-\, \rightarrow\, p_{\pm{1\over 2}}
\,+\,\overline\Delta_{\mp{1\over 2}}$.  Here,  the physically dominant degree
of freedom is {\it not}  $\Delta_{\pm{3\over 2}}$ but  $\Delta_{\pm{1\over
2}}$ --- the degree of freedom which for massless case identically vanishes.

\section{Generalised Melosh Transformation -- 
The Connection with the Instant Form}
 
In a representation appropriate for comparison with the front form
spinors ${\cal U}_h(\lowercase {p}^\mu)$ and ${\cal
V}_h(\lowercase{p}^\mu)$, the instant form hadronic spinors
$u_\sigma(\,{\underline p}^\mu)$ and $v_\sigma (\,{\underline
p}^\mu)$, $\sigma\eq j,\,j\,-\,1,\ldots,-\,j$, were recently
constructed explicitly (following Weinberg \cite{Weinberg} and Ryder
\cite{Ryder}) in Refs. [3(a-c), 3(g)]. A brief report, sufficient for
the present discussion, can be found in Ref. [3(b)]. Here we only
remark that the construction of instant form spinors follows the steps
outlined in Eqs.(\ref{mikolaj}-\ref{ml}) above, with the only
difference that one starts with transformation $\Lambda$ of
Eq.(\ref{il}) rather than $L$ of Eq.(\ref{mikolaj}).

The instant form hadronic spinors
of Refs.
[3(a-c), 3(g)] satisfy the following normalisation properties
\begin{eqnarray}
{\overline u}_\sigma (\,\underline {p}^\mu)\,\,{u}_{\sigma'}
(\underline{p}^\mu) &\eq&
m^{2\,j}\,\delta_{\sigma \sigma'} \\
{\overline v}_\sigma (\,\underline{p}^\mu)\,\,v_{\sigma'}(\,\underline{p}^\mu) &\eq&
-\,m^{2\,j}\,\delta_{\sigma\sigma'} \\
{\overline u}_\sigma (\,\underline{p}^\mu)\,\,{v}_{\sigma'}(\,\underline{p}^\mu) &\eq& 0\eq
{\overline v}_\sigma (\,\underline{p}^\mu)\,\,{u}_{\sigma'}(\,\underline{p}^\mu) \,\,.
\end{eqnarray}
where
\beq
\overline{u}_\sigma(\,\underline{p}^\mu) \eq  \left[{u}_\sigma(\,\underline{p}^\mu)
\right]^\dagger\gamma^\circ,~~
\overline{v}_\sigma(\,\underline{p}^\mu) \eq  \left[{v}_\sigma(\,\underline{p}^\mu)
\right]^\dagger\gamma^\circ
\eeq
with $\gamma^\circ$ having the form identical to $\Gamma^\circ$ of
Eq.(\ref{go}). 

In what follows we assume that $p^\mu$ and $\underline{p}^\mu$
correspond to the same physical momentum.  The connection between the
front form and instant form spinors is then established by noting that
on general algebraic grounds we can express the instant form hadronic
spinors as linear combination of the front form hadronic spinors. That
is
\begin{eqnarray}
u_\sigma( \,\underline{p}^\mu)&\eq& \Omega^{(u\,{\cal U})}_{\sigma h}\,{\cal
U}_h( p^\mu) \,+\, \Omega^{(u\,{\cal V})}_{\sigma h}\,{\cal V}_h(p^\mu) \,,
\label{one} \\
v_\sigma( \,\underline{p}^\mu)&\eq& \Omega^{(v\,{\cal U})}_{\sigma h}\,{\cal
U}_h( p^\mu) \,+\, \Omega^{(v\,{\cal V})}_{\sigma h}\,{\cal V}_h(p^\mu) \,,
\label{two}
\end{eqnarray}
where the sum on the repeated indices is implicit.

We now multiply Eq. (\ref{one}) by ${\overline{\cal U}}_{h'}(p^\mu)$ from the left,
and using the orthonormality relations, Eqs. (\ref{three}-\ref{five}),
we get
\beq
\Omega^{(u\,{\cal U})}_{\sigma h} \eq {1\over m^{2j}}\left[\,{\overline{\cal U}}_h
(p^\mu)\,
u_\sigma(\,\underline{p}^\mu)\right]\,\,. \eeq
Similarly by multiplying Eq. (\ref{two}) from the left by
${\overline{\cal V}}_{h'}(p^\mu)$ and again using
the orthonormality relations, Eqs. (\ref{three}-\ref{five}), we obtain
\beq
\Omega^{(v\,{\cal V})}_{\sigma h} \eq -\,{1\over m^{2j}}\left[\,
{\overline{\cal V}}_h
(p^\mu)\,
v_\sigma(\,\underline{p}^\mu)\right]\,\,.
\eeq
Further it is readily verified, e.g. by using the results presented in Appendix
B here and explicit expressions for $u_\sigma(\,\underline{p}^\mu)$ and
$v_\sigma(\,\underline{p}^\mu)$ found in Refs. [3(a-c),3(g)], that
\beq
{\overline{\cal V}}_h(p^\mu)\,u_\sigma(\,\underline{p}^\mu)\eq 0
\eq
{\overline{\cal U}}_h(p^\mu)\,v_\sigma(\,\underline{p}^\mu);
\eeq
which yields
\beq
\Omega^{(u\,{\cal V})}_{\sigma h} \eq 0 \eq
\Omega^{(v\,{\cal U})}_{\sigma h}\,\, .
\eeq
Finally, we exploit the facts
\beq
\left\{\Gamma^5,\,\Gamma^\circ\right\}\eq 0,~~{\Gamma^5}^\dagger \eq
\Gamma^5,~~ {\rm and}~\left(\Gamma^5\right)^2\eq I,
\eeq
to conclude that ${\overline {\cal
V}}_h(p^\mu)\,v_\sigma(\,\underline{p}^\mu)
\eq -\, {\overline {\cal U}}_h( p^\mu)\,u_\sigma(\,\underline{p}^\mu)$. 
Thus the matrix which connects the instant form spinors with front 
form spinors reads
\beq
\Omega(j)\eq
\left[
\begin{array}{ccc}
B(j)&{~}& 0 \\ \\
0 &{~}& B(j)
\end{array}
\right]\,\,,\label{six}
\eeq
where $B(j)$ is a $(2j\,+\,1) \times(2j\,+\,1)$ matrix with elements
$B_{\sigma h}\eq
\overline{\cal U}_h(p^\mu)\,u_\sigma(\,\underline{p}^\mu)\eq
\Omega^{(u{\cal U})}_{\sigma h} \eq
\Omega^{(v{\cal V})}_{\sigma h}$.
 
The explicit expressions for $\Omega(j)$ are presented in Appendix C. For spin
one half the transformation matrix $\Omega(1/2)$ computed by us coincides with
the celebrated ``Melosh transformation'' given by Melosh in Ref. [5, Eq.  26] 
and by Dziembowski in Ref.  [6, Eq. A8]. As
formally demonstrated by Kondratyuk and Terent'ev \cite{KT}, the transformation
matrix $\Omega({j})$ represents a pure rotation of the spin basis. 
However, since $\Omega(j)$ has block zeros off-diagonal, what manifestly
emerges here is that this rotation does not mix the even and odd spinor
parity spinors.

\section{Summary}
 
Within the framework of Weinberg-Soper formalism \cite{Weinberg,Soper} we
constructed explicit hadronic spinors for arbitrary spin in the front form, and
established their connection with the instant form fields. For a given a spin,
$j$, there are $(2j\,+\,1)$ hadronic spinors with {\it even} spinor  parity,
${\cal U}_h(p^\mu)$; and $(2j\,+\,1)$ hadronic spinors with {\it odd} spinor
parity, ${\cal V}_h(p^\mu)$.  The normalisation of these spinors 
is so chosen that for the {\it massless} particles the ${\cal U}_h(\po^\mu)$
and ${\cal V}_h(\po^\mu)$  identically vanish; and only ${\cal U}_{h\eq\pm
j}(p^\mu)$ and ${\cal V}_{h\eq\pm j}(p^\mu)$ survive. The simplest choice of
this normalisation is given by Eq. (\ref{norm}). Next we constructed the matrix
$\Omega(j)$ which provides the connection between the front form hadronic
spinors with the more familiar [i.e. more ``familiar'' at least for spin one
half case] instant form hadronic spinors. We verified that  the transformation
matrix $\Omega(1/2)$ coincides with the well known ``Melosh transformation''
\cite{Melosh,Dziembowski}, and the spin one half spinors are in
agreement with the previous results of Lepage and Brodsky \cite{lepage}.
Explicit results for ${\cal U}_h(p^\mu)$, ${\cal V}_h(p^\mu)$ and $\Omega(j)$
up to spin two are found in Appendixes B and C here.

\acknowledgements
 
Stan Brodsky and Terry Goldman helped us clear the discussion of the massless
case and the high momentum limit for massive case; for this we thank them.
DVA extends his thanks to Ovid C. Jacob, Mikkel B. Johnson and R. M.  Thaler
for insightful conversations on the general subject of this work. He 
acknowledges a indebetedness to Dave  Ernst for our previous collaborative
work on this subject, and the insights which are carried over into the present
work. He also thankfully acknowledges financial support via a postdoctoral
fellowship by the Los Alamos National Laboratory.  Part of this work was done
while one of the authors (MS) was a visiting faculty at the Department of
Physics of the Texas A\&M University. He would like to thank
Professor David J. Ernst and members of the Department of Physics for warm
hospitality extended to him during his stay.

\widetext
\appendix{Expansions for the $\exp\left(\eta\,\lowercase{\hat b}\cdot\vec
J\,\right)$ and $ \exp\left(-\,\eta\,\lowercase{\hat b}^\ast\cdot\vec
J,\right)$ up to Spin Two}
\footnotesize
\widetext
This appendix provides the expansions for the $\exp\left(\eta\,\hat b\cdot\vec
J\,\right)$ and $ \exp\left(-\,\eta\,\hat b^\ast\cdot\vec J\,\right)$, up to
$j\eq 2$,  which appear in the spinor-boost matrix $M(L)$, Eq.
(\ref{ml}).
 
\centerline{ $j\eq{1\over 2}$ }
\begin{eqnarray}
\exp\left(\eta\,\hat b\cdot\vec J\,\right)
&\eq&
\cosh\left({\eta/ 2}\right) \,I\,+\,
\left(\hat b\cdot\vec\sigma\right)\,\sinh\left({\eta/ 2}\right)\,, \label{aa}
\\
\exp\left(-\,\eta\,\hat b^\ast\cdot\vec J\,\right)
&\eq&
\cosh\left({\eta/2}\right)\,I \,-\,
\left(\hat b^\ast\cdot\vec\sigma\right)\,\sinh\left({\eta/ 2}\right)\,.
\label{ab}
\end{eqnarray}
\def\bj{\left(\hat b\cdot\vec J\,\right)}
\def\bsj{\left(\hat b^\ast\cdot\vec J\,\right)}
\centerline {$j\eq{1}$ }
\begin{eqnarray}
\exp\left(\eta\,\hat b\cdot\vec J\,\right)
&\eq&
I\,+\,2\bj^2\sinh^2(\eta/2)\,+\,2\bj\cosh(\eta/2)\,\sinh(\eta/2)\,,\\
\exp\left(-\,\eta\,\hat b^\ast\cdot\vec J\,\right)
&\eq&
I\,+\,2\bsj^2\sinh^2(\eta/2)\,-\,2\bsj\cosh(\eta/2)\,\sinh(\eta/2)\,.
\end{eqnarray}
 
\def\bbj{\left(2\hat b\cdot\vec J\,\right)}
\def\bbsj{\left(2\hat b^\ast\cdot\vec J\,\right)}
\def\e{(\eta/2)}
\centerline{ $j\eq{3\over 2}$ }
\begin{eqnarray}
\exp\left(\eta\,\hat b\cdot\vec J\,\right)
\eq&&
\cosh\e\left[I\,+\,{1\over 2}\left\{\bbj^2\,-\,I\right\}\,\sinh^2\e\right]
\nonumber\\
&&{}\,+\,\bbj\sinh\e\left[I\,+\,{1\over 6}\left\{\bbj^2\,-\,I\right\}\sinh^2\e
\right] \,,\\
\exp\left(-\,\eta\,\hat b^\ast\cdot\vec J\,\right)
\eq&&
\cosh\e\left[I\,+\,{1\over 2}\left\{\bbsj^2\,-\,I\right\}\,\sinh^2\e\right]
\nonumber\\
&&{}\,-\,\bbsj\sinh\e\left[I\,+\,{1\over 6}\left\{\bbsj^2\,-\,I\right\}\sinh^2\e
\right]\,.
\end{eqnarray}
 
\centerline{ $j\eq 2$ }
{\footnotesize
\begin{eqnarray}\hskip-.5in
\exp\left(\eta\,\hat b\cdot\vec J\,\right)
\eq&&
I\,+\,2\bj^2\sinh^2\e\,+\,{2\over 3}\bj^2\left\{\bj^2\,-I\right\}\sinh^4\e
 \nonumber\\
&&{}\,+\, 2\bj\cosh\e\sinh\e\,+\,{4\over
3}\bj\left\{\bj^2\,-\,I\right\}\cosh\e\sinh ^3\e\,,\\
\exp\left(-\eta\,\hat b^\ast\cdot\vec J\,\right)
\eq&&
I\,+\,2\bsj^2\sinh^2\e\,+\,{2\over 3}\bsj^2\left\{\bsj^2\,-I\right\}
\sinh^4\e \nonumber \\
&&{}\,-\, 2\bsj\cosh\e\sinh\e\,-\,{4\over 3}\bsj\left\{\bsj^2\,-\,I\right\}
\cosh\e\sinh
^3\e\,.
\end{eqnarray}
}
In Eqs. (\ref{aa},\ref{ab}) the $\vec \sigma$ are the standard
\cite{Ryder} Pauli matrices.
In this appendix, $I$ are  the $(2j+1)\times(2j+1)$ identity matrices.

\appendix{Front Form Hadronic Spinors up to Spin two}
\footnotesize
We begin with  collecting together front form hadronic spinors up to
spin 2 {\it for even spinor parity}, first. In what follows we use the notation
$\pr\eq p_x\,+\,i\,p_y$ and $\pl\eq p_x\,-\,i\,p_y$, c.f. Eq.(\ref{bs}),
\vskip 0.3in\noindent
Spin one half hadronic spinors with
even spinor parity:
 
\def\ca{{1\over{2}}\sqrt{1\over \pp}}
\beq
{\cal U}_{+\,{1\over 2}}(p^\mu)\eq\ca\,\left[
\begin{array}{c}
\pp\,+\,m\\ \\
\pr\\ \\
\pp\,-\,m\\ \\
\pr
\end{array}
\right]
\,,~~
{\cal U}_{-\,{1\over 2}}(p^\mu)\eq\ca\left[
\begin{array}{c}
-\,\pl\\ \\
\pp\,+\,m\\ \\
\pl\\ \\
-\,\pp\,+\,m
\end{array}
\right]
\,.\label{ua}
\eeq
 
\vskip 0.3in\noindent
Spin one  hadronic spinors with
even spinor parity:
{\footnotesize\hskip-.5in
\beq
{\cal U}_{+1}(p^\mu)\eq
{1\over 2}\,\left[
\begin{array}{c}
\pp\,+\,\left(m^2/\pp\right)\\ \\
{\sqrt{2}}\,\pr \\ \\
{\pr^2/\pp}\\ \\
\pp\,-\,\left({m^2/\pp} \right)\\ \\
{\sqrt{2}} \,\pr\\ \\
{\pr^2/\,\pp}
\end{array}
\right],\,~~
{\cal U}_{0}(p^\mu)\eq
m\,{\sqrt{1\over 2}}\,\left[
\begin{array}{c}
-\,{\pl/\pp} \\\\
\sqrt {2} \\ \\
\,{\pr/\pp} \\ \\
\,{\pl/\pp} \\\\
0\\ \\
\,{\pr/\pp}
\end{array}
\right]\,~~
{\cal U}_{-1}(p^\mu)\eq
{1\over 2}\,\left[
\begin{array}{c}
{\pl^2/\pp}  \\ \\
-\,{\sqrt{2}} \,\pl\\ \\
\pp\,+\,\left({m^2/\pp} \right)\\ \\
-\,{\pl^2/\pp}  \\ \\
{\sqrt{2}}\, \pl\\ \\
-\,\pp\,+\,\left({m^2/\pp} \right)
\end{array}
\right]\,.\label{ub}
\eeq
}

\def\t{\sqrt{3}}
 
\vskip 0.3in\noindent
Spin three half  hadronic spinors with
even spinor parity:
\beq
{\cal U}_{+{3\over 2}}(p^\mu)\eq
{{1\over 2}\,\sqrt{1\over\pp}}\left[
\begin{array}{c}
{\pp}^2\,+\,\left(m^3/{\pp} \right) \\ \\
\sqrt{3}\, \pr\,{\pp} \\ \\
\t\,\pr^2 \\ \\
\pr^3/{\pp} \\ \\
{\pp}^2\,-\,\left(m^3/{\pp}\right) \\ \\
\sqrt{3}\, \pr\,{\pp} \\ \\
\t\,\pr^2 \\ \\
\pr^3/{\pp}
\end{array}
\right]\,,~~
{\cal U}_{+{1\over 2}}(p^\mu)\eq
{{m\over 2}\,\sqrt{1\over\pp}}\left[
\begin{array}{c}
-\,\t\,m\,\pl/\pp \\ \\
\pp\, +\,m\\ \\
2\,\pr \\ \\
\t\,{\pr}^2/\pp \\ \\
\t\,m\,\pl/\pp \\ \\
\pp\,-\,m \\ \\
2\,\pr \\ \\
\t\,\pr^2/\pp
\end{array}
\right]\,,\label{uca}
\eeq
 
\beq
{\cal U}_{-{1\over 2}}(p^\mu)\eq
{{m\over 2}\,\sqrt{1\over\pp}}\left[
\begin{array}{c}
\t\,\pl^2/\pp \\ \\
-\,2\,\pl \\ \\
\pp\,+\,m \\ \\
\t\,m\,\pr/\pp \\ \\
-\,\t\,\pl^2/\pp \\ \\
2\,\pl \\ \\
-\,\pp\,+\,m \\ \\
\t\,m\,\pr/\pp
\end{array}
\right]\,,~~
{\cal U}_{-{3\over 2}}(p^\mu)\eq
{{1\over 2}\,\sqrt{1\over\pp}}\left[
\begin{array}{c}
-\,\pl^3/\pp \\ \\
\t\,\pl^2 \\ \\
-\,\t\,\pl\,\pp \\ \\
{\pp}^2\,+\,\left(m^3/\pp\right) \\ \\
\pl^3/\pp \\ \\
-\t\,\pl^2 \\ \\
\t\,\pl\,\pp \\ \\
-\,{\pp}^2\,+\,\left(m^3/\pp \right)
\end{array}
\right]\,\,.\label{ucb}
\eeq
 
\newpage
\noindent
Spin two hadronic spinors with
even spinor parity:
\def\s{\sqrt{6}}
\beq
{\cal U}_{+{2}}(p^\mu)\eq
{1\over 2}\,\left[
\begin{array}{c}
{{\pp}^2}\,+\,\left(m^4/{\pp}^2\right) \\ \\
2\,\pr\,\pp \\ \\
\s\,\pr^2 \\ \\
2\,{\pr}^3/\pp \\ \\
\pr^4/{\pp}^2 \\ \\
{\pp}^2\,-\,\left(m^4/{\pp}^2\right) \\ \\
2\,\pr\,\pp \\ \\
\s\,\pr^2 \\ \\
2\,\pr^3/\pp \\ \\
\pr^4/{\pp}^2
\end{array}
\right]\,,~~
{\cal U}_{+{1}}(p^\mu)\eq
{m\over 2}\,\left[
\begin{array}{c}
-\,2\,m^2\,\pl/{\pp}^2 \\ \\
\pp\,+\,\left(m^2/\pp\right) \\ \\
\s\,\pr \\ \\
3\,\pr^2/\pp \\ \\
2\,\pr^3/{\pp}^2 \\ \\
2\,m^2\,\pl/{\pp}^2 \\ \\
\pp\,-\,\left(m^2/\pp\right) \\ \\
\s\,\pr \\ \\
3\,\pr^2/\pp \\ \\
2\,\pr^3/{\pp}^2
\end{array}
\right]\,,\label{uda}
\eeq

\beq
{\cal U}_{{0}}(p^\mu)\eq
{m^2\over 2}\,\left[
\begin{array}{c}
\s\,\pl^2/{\pp}^2 \\ \\
-\,\s\,\pl/\pp \\ \\
2 \\ \\
\s\,\pr/\pp \\ \\
\s\,\pr^2/{\pp}^2 \\ \\
-\,\s\,\pl^2/{\pp}^2 \\ \\
\s\,\pl/\pp \\ \\
0 \\ \\
\s\,\pr/\pp \\ \\
\s\,{\pr}^2/{\pp}^2
\end{array}
\right]\,,~~
{\cal U}_{-{1}}(p^\mu)\eq
{m\over 2}\,\left[
\begin{array}{c}
-\,2\,\pl^3/{\pp}^2 \\ \\
3\,\pl^2/\pp \\ \\
-\,\s\,\pl \\ \\
\pp\,+\,\left(m^2/\pp\right) \\ \\
2\,m^2\,\pr/{\pp}^2 \\ \\
2\,\pl^3/{\pp}^2 \\ \\
-\,3\,\pl^2/\pp \\ \\
\s\,\pl \\ \\
-\,\pp\,+\,\left(m^2/\pp\right) \\ \\
2\,m^2\,\pr/{\pp}^2 \\ \\
\end{array}\right]\, ,\label{udb}
\eeq
 
\beq
{\cal U}_{-{2}}(p^\mu)\eq
{1\over 2}\,\left[
\begin{array}{c}
\pl^4/{\pp}^2 \\ \\
-\,2\,\pl^3/\pp \\ \\
\s\,\pl^2 \\ \\
-\,2\,\pl\,\pp \\ \\
{\pp}^2\,+\,\left(m^4/{\pp}^2\right) \\ \\
-\,\pl^4/{\pp}^2 \\ \\
2\,\pl^3/\pp \\ \\
-\,\s\,\pl^2 \\ \\
2\,\pl \,\pp \\ \\
-\,{\pp}^2\,+\,\left(m^4/{\pp}^2\right)
\end{array}
\right]\,\,.\label{udc}
\eeq
 
An examination of the spinor-boost matrix $M(L)$,
Eq.(\ref{ml}), implies that the {\it odd } spinor parity hadronic
spinor can be obtained from the hadronic spinor of the {\it
even} spinor parity via the following simple relation 
\beq 
{\cal V}_h(p^\mu)\eq \Gamma^5 \,{\cal U}_h(p^\mu)\,\,,
\eeq
where the matrix $\Gamma^5$ defined in Eq.(\ref{go}) interchanges the top $(2j\,+\,1)$ elements with the
bottom $(2j\,+\,1)$ elements of the hadronic spinors.

\appendix{The Explicit Expressions for $\Omega(\lowercase{j})$, the Generalised
Melosh Transformation, up to Spin Two}
\footnotesize

In this appendix we present explicit expressions for the matrix $\Omega(j)$,
Eq. (\ref{six}),
which connects the front form hadronic spinors with the instant form hadronic
spinors via Eqs. (\ref{one},\ref{two}).
As in Appendix B,
$\pr\eq p_x\,+\,i\,p_y$ and $\pl\eq p_x\,-\,i\,p_y$, in what follows.
\vskip 0.3in\noindent
 
For spin one half,  the matrix connecting the instant form hadronic spinors with the front form spinors is
\def\hc{ {1\over {\left[2\,\l E\,+\,m\r\,\pp\,\right]^{1/2} }}  }
\beq
\Omega\l 1/2 \r \eq \hc \,
\left[
\begin{array}{cccc}
\beta(1/2)&{}&{}&0 \\ \\
0&{}&{}&\beta(1/2)
\end{array}
\right]\,,\label{ha}
\eeq
where  the $2\times2$ block matrix $\beta(1/2)$ is defined as
\beq
\beta\l 1/2\r\eq
\left[
\begin{array}{ccc}
\pp\,+\,m&{}&-\,\pr \\ \\
\pl&{}& \pp\,+\,m
\end{array}
\right]\,.\label{hb}
\eeq
 
 
For spin one, the matrix connecting the instant form hadronic spinors
with the front form spinors is
\def\oc{ {1\over {\left[2\,\l E\,+\,m\r\,\pp\,\right] }}  }
\beq
\Omega\l 1 \r \eq \oc \,
\left[
\begin{array}{ccc}
\beta(1)&{}&0 \\ \\
0&{}&\beta(1)
\end{array}
\right]\,,\label{oa}
\eeq
where  the $3\times3$ block matrix $\beta(1)$ is defined as
\beq
\beta\l 1 \r \eq
\left[
\begin{array}{ccccc}
\l\pp\,+\,m\r^2 &{}& -\,\sqrt{2}\,\l\pp\,+\,m\r\,\pr &{~~}& \pr^2 \\ \\
\sqrt{2}\,\l\pp\,+\,m\r\,\pl &{~~}&2\,\left[\l E\,+\,m\r\,\pp\,-\,\pr\pl\right]
&{~~}&-\,\sqrt{2}\,\l\pp\,+\,m\r\,\pr \\ \\
\pl^2 &{~~}&\sqrt{2}\,\l\pp\,+\,m\r\,\pl &{~~}& \l\pp\,+\,m\r^2
\end{array}
\right]\,\,.\label{ob}
\eeq
 
 
For spin three half,  the matrix connecting the instant form hadronic spinors 
with the front form spinors is
\def\tc{ {1\over {\left[2\,\l E\,+\,m\r\,\pp\,\right]^{3/2} }}  }
\beq
\Omega\l 3/2 \r \eq \tc \,
\left[
\begin{array}{cccc}
\beta(3/2)&{}&{}&0 \\ \\
0&{}&{}&\beta(3/2)
\end{array}
\right]\,,\label{ta}
\eeq
where  the $4\times4$ block matrix $\beta(3/2)$ is defined as
\newpage
{
\FL
\begin{eqnarray}
&&\beta\l 3/2\r\eq \\ \nonumber
&&\left[
\begin{array}{ccccccc}
\l\pp\,+\,m\r^3 &{~}& -\,\sqrt{3}\,\l\pp\,+\,m\r^2\,\pr &{~}&
\sqrt{3}\,\l\pp\,+\,m\r\,\pr^2 &{~}& -\,\pr^3 \\ \\
\,\sqrt{3}\,\l\pp\,+\,m\r^2\,\pl &{~}& \left[\l\pp\,+\,m\r^2\,-\,
2\,\pr\,\pl\right]\,\l\pp\,+\,m\r &{~}&
-\left[2\,\l\pp\,+\,m\r^2-\pr\,\pl\right]\,\pr &{~}&
\sqrt{3}\,\l\pp\,+\,m\r\,\pr^2  \\ \\
\sqrt{3}\,\l\pp\,+\,m\r\,\pl^2  &{~}&
\left[2\,\l\pp\,+\,m\r^2-\pr\,\pl\right]\,\pl &{~}&
\left[\l\pp\,+\,m\r^2\,-\,
2\,\pr\,\pl\right]\,\l\pp\,+\,m\r &{~}&
-\,\sqrt{3}\,\l\pp\,+\,m\r^2\,\pr \\ \\
\pl^3 &{~}& \sqrt{3}\,\l\pp\,+\,m\r\,\pl^2  &{~}&
\sqrt{3}\,\l\pp\,+\,m\r^2\,\pl&{~}& \l\pp\,+\,m\r^3
\end{array}
\right]\,.\label{tb}
\end{eqnarray}
}

\def\c{{1\over {\left[2\,\l E\,+\,m\r\,p^+\,\right]^2}} }
\def\aa{\left(p^+\,+\,m\right)^4}
\def\ba{2\,\left({\pp}\,+\,m\right)^3\,\pl}
\def\ca{\sqrt{6}\,\left({\pp}\,+\,m\right)^2\,\pl^2}
\def\da{2\,\left({\pp}\,+\,m\right)\,\pl^3}
\def\ea{\pl^4}
\def\ab{-\,2\,\l{\pp}\,+\,m\r^3\,\pr}
\def\bb{2\left[\l E\,+\,m\r\,{\pp}\,-\,2\,\pr\,\pl\right]\,\l{\pp}\,+\,m\r^2}
\def\cb{\sqrt{6}\,\left[\l
E\,+\,m\r\,{\pp}\,-\,\pr\,\pl\right]\,\l{\pp}\,+\,m\r\,\pl}
\def\db{\left[6\,{\pp}\l E\,+\,m\r\,-\,4\,\pr\,\pl\right]\,\pl^2}
\def\eb{2\,\l{\pp}\,+\,m\r\,\pl^3}
\def\ac{\sqrt{6}\l{\pp}\,+\,m\r^2\,\pr^2}
\def\bc{\sqrt{6}\,\left[-\,\l E\,+\,m\r\,{\pp}\,+\,\pr\,\pl\right]
\l{\pp}\,+\,m\r\,\pr}
\def\cc{2\left[2\,{\pp}^2\,\l
E\,+\,m\r^2\,-\,3\,\l{\pp}\,+\,m\r^2\,\pr\,\pl\right]}
\def\dc{\sqrt{6}\left[\l E\,+\,m\r\,{\pp}\,-\,\pr\,\pl\right]\,\l{\pp}\,+\,m\r\,
\pl}
\def\ec{\sqrt{6}\,\left({\pp}\,+\,m\right)^2\,\pl^2}
\def\ad{-\,2\,\l{\pp}\,+\,m\r\,\pr^3}
\def\bd{\left[6\,{\pp}\,\l E\,+\,m\r\,-\,4\,\pr\,\pl\right]\,\pr^2}
\def\cd{\sqrt{6}\,\left[-\,\l E\,+\,
m\r\,{\pp}\,+\,\pr\,\pl\right]\,\l{\pp}\,+ \,m\r\,\pr}
\def\dd{2\left[\l E\,+\,m\r\,{\pp}\,-\,2\,\pr\,\pl\right]\,\l{\pp}\,+\,m\r^2}
\def\ed{2\,\l{\pp}\,+\,m\r^3\,\pl}
\def\ae{\pr^4}
\def\be{-\,2\l{\pp}\,+\,m\r\,\pr^3}
\def\ce{\sqrt{6}\,\l{\pp}\,+\,m\r^2\,\pr^2}
\def\de{-\,2\,\l{\pp}\,+\,m\r^3\pr}
\def\ee{\l{\pp}\,+\,m\r^4}
 
For spin two, the matrix connecting the instant form hadronic spinors
with the front form spinors is
\beq
\Omega\l 2 \r \eq \c \,
\left[
\begin{array}{ccc}
\beta(2)&{}&0 \\ \\
0&{}&\beta(2)
\end{array}
\right]\,,\label{twoa}
\eeq
where  the $5\times5$ block matrix $\beta(2)$ is defined via the following five
columns
\begin{eqnarray}
&&\beta(2)_{\alpha,1}\eq
\left[
\begin{array}{c}
\aa \\ \\
\ba \\ \\
\ca \\ \\
\da \\ \\
\ea
\end{array}
\right]\,,~~
\beta(2)_{\alpha,2}\eq
\left[
\begin{array}{c}
\ab \\ \\
\bb \\ \\
\cb \\ \\
\db \\ \\
\eb
\end{array}
\right]\,, \nonumber\\\nonumber \\
&& \beta(2)_{\alpha,3}\eq
\left[
\begin{array}{c}
\ac \\ \\
\bc \\ \\
\cc \\ \\
\dc \\ \\
\ec
\end{array}
\right]\,,\nonumber \\\nonumber \\
&& \beta(2)_{\alpha,4}\eq
\left[
\begin{array}{c}
\ad \\ \\
\bd \\ \\
\cd \\ \\
\dd \\ \\
\ed
\end{array}
\right]\,,~~
\beta(2)_{\alpha,5}\eq
\left[
\begin{array}{c}
\ae \\ \\
\be \\ \\
\ce \\ \\
\de \\ \\
\ee
\end{array}
\right]\,.\label{twob}
\end{eqnarray}

\mediumtext
\begin{table}
\caption{Algebra associated with the stability group of the  $x^+\eq 0$ plane.
The Commutator  $\big[\,$Element in the {\it first} column,
Element in the {\it first} row$\,\big]\eq$The element at the
intersetction of {\it the} row and column.}
\begin{tabular}{clllllll}
\multicolumn{1}{c}{ } &\multicolumn{1}{c}{$P_1$}
&\multicolumn{1}{c}{$P_2$} &\multicolumn{1}{c}{$J_3$}
&\multicolumn{1}{c}{$K_3$} &\multicolumn{1}{c}{$P_{-}$}
&\multicolumn{1}{c}{$G_1$} &\multicolumn{1}{c}{$G_2$}\\
\tableline
 $P_1$ & $0$ & $0$ & $-iP_2$ & $0$ & $0$ & $iP_-$ & $0$\\
 $P_2$ & $0$ & $0$ & $iP_1$ & $0$ & $0$ & $0$ & $iP_-$\\
 $J_3$ & $iP_2$ & $-iP_1$ & $0$ & $0$ & $0$ & $iG_2$ & $-iG_1$\\
 $K_3$ & $0$ & $0$ & $0$ & $0$ & $iP_-$ & $iG_1$ & $iG_2$\\
 $P_-$ & $0$ & $0$ & $0$ & $-iP_-$ & $0$ & $0$ & $0$\\
 $G_1$ & $-iP_-$ & $0$ & $-iG_2$ & $-iG_1$ & $0$ & $0$ & $0$\\
 $G_2$ & $0$ & $-iP_-$ & $iG_1$ & $-iG_2$ & $0$ & $0$ & $0$\\
\end{tabular}
\end{table}

\end{document}